\documentclass[twocolumn,preprint,showpacs,preprintnumbers,amsmath,amssymb]{revtex4}

\usepackage{graphicx}
\usepackage{dcolumn}
\usepackage{bm}

\begin{document}

\preprint{APS/123-QED}

\title{Experimental balance energies and 
 isospin-dependent nucleon-nucleon cross-sections\\}
\author{Sanjeev Kumar}
\author{Rajni}
\author{Suneel Kumar}%
 \email{suneel.kumar@thapar.edu}
\affiliation{%
School of Physics and Materials Science, Thapar University, Patiala-147004, Punjab (India)\\
}%

%

\date{\today}

\begin{abstract}
The effect of different isospin-dependent cross-section on directed flow is studied 
for variety of systems(for which experimental balance energies are available) using
an isospin-dependent Quantum Molecular Dynamic (IQMD) model. We show that 
balance energies are sensitive towards isospin-dependent  
cross-sections for light systems, while nearly no effect exist for heavier nuclei. A reduced cross-section 
 $\sigma=0.9 \sigma_{NN}$ with stiff equation of state is able to explain experimental 
balance energies in most of systems. A power law behaviour is also given for the mass dependence of balance energy,
which also follow N/Z dependence.    	  
\end{abstract}

\pacs{25.70.-z, 25.75.Ld}
\maketitle
\section{Introduction}
The heavy-ion physics branch has been renewed interest very recently. This ranges from the fusion probabilities 
\cite{Puri92} to symmetry energy dependence at intermediate energies as well as fragmentation of colliding matter
\cite{Singh00}.  
 One observable that has been used extensively 
 for extracting information from heavy-ion collisions is the collective in-plane flow of various
particles \cite{Pan93,Sorge97,Ollitrauet98,Westfall93,Pak96,Pak97,Li96,Sood04,Leiwen98}. 
Apart from transverse in-plane flow, one has also proposed, e.g differential \cite{Li99}
and elliptical flow \cite{Zhang99} etc.\\
In general, collective flow in heavy-ion collisions is 
affected by both the nuclear mean field potential and nucleon-nucleon (NN) cross-sections. One should 
also keep in the mind that reaction dynamics depends also on the incident energy as well as on the 
impact parameter of the reaction \cite{Westfall93,Pak96,Pak97,Li96,Sood04}.
 At low incident energies, reaction dynamics is 
dominated by the attractive nuclear mean field potential which results in deflection to negative angle.
 Worth mention, at these enegies the phenomenon fusion, 
and cluster radioactivity are dominated \cite{RGupta93,Puri92}. With increasing incident energy, 
repulsive nucleon-nucleon scattering becomes important
and results in reduced negative flow caused by the attractive mean field potential. 
As a result, at a certain incident energy, called the balance energy, the in-plane flow vanishes
as a result of cancellation between these two competing effects [10-14]. The composite dependence of 
the $E_{bal}$ on the mean field and nucleon-nucleon cross-sections ($\sigma_{NN}$) can be sorted out by 
noticing the sensitivity of $E_{bal}$ on the system size, impact parameter as well as isospin degree 
of freedom of the reaction [6-10].\\
 Experimentally, balance energy is observed for different systems ranging from 
$^{12}C+^{12}C$ to $^{197}Au+^{197}Au$ \cite{Westfall93,Pak96,Pak97,Sullivain90,Cussol02}. 
The very accurate measurement of the $E_{bal}$ in $^{197}Au+^{197}Au$ \cite{Cussol02} has generated a
renewed interest in the field. Unfortunately, these studies have not provided any significant 
contribution of isospin effects towards the balance energy. Later on, Pak {et al.} \cite{Pak97} demonstrated 
the isospin effect on the collective flow and balance energy at central and peripheral geometries. 
These findings were limited only for $^{58}Fe+^{58}Fe$ and $^{58}Ni+^{58}Ni$ systems. Theoretically, 
the disappearance of directed flow is studied using the Boltzmann Uehling Uehlenbeck (BUU) model. 
\cite{Westfall93,Pak96,Li99,Sullivain90} and Quantum Molecular Dynamics (QMD) model [3-5,10-21]. Different 
theoretical attempts considered either a stiff or soft equation of state along with a variety of 
NN cross-sections. Very recently, Puri and co-workers \cite{Sood04}, conducted a very detailed 
analysis on the balance energy over entire periodic table with masses between 24 and 394. These 
study shed light on various aspect of nuclear dynamics. Unfortunately, this study along with 
all other studies reported in literature are limited to Central/Semi-Central Collisions only \cite{Sood04}.
Following this work, the detailed analysis on the the semi-central and peripheral collision 
is performed by the Puri and co-workers in 2010 \cite{Chugh10}. All these studies indicated enhanced 
cross-section of 40-55mb with stiff equation of state to verify the balance energy in intermediate
energy heavy-ion collisions. All these studies were independent of isospin effects. The first study showing
the isospin effects on the collective flow and balance energy was reported by Li {et al.}\cite{Li96}
 using the isospin dependent Boltzmann Uehling Uehlenbeck (IBUU) model, where strong dependence of 
isospin effects was observed. In another contribution
\cite{Zhang99}, they suggested the demand of reduced isospin dependent cross-section($\sigma = 0.88\sigma_{NN}$)
to better explain the experimental data. Chen {et al.}\cite{Leiwen98} studied the effect of 
isospin degree of freedom on the balance energy using isospin dependent Quantum Molecular Dynamics 
(IQMD) model, which was an improved version of 
original QMD model \cite{Sood04,Aichelin91,Singh00}. The calculated results were found to differ from 
the data at all colliding geometries. Recently, Gautam {et al.} \cite{Gautam10} also studied the 
isospin effect on the balance energy by using IQMD model \cite{Hartnack98}, and also compared their findings
with the other theoretical and experimental findings. They demanded to take care the Gaussian width(L) 
and cross-section ($\sigma = 0.88\sigma_{NN}$), while studying the isospin effects in 
intermediate energy heavy-ion collisions.\\
From the above, it is cleared that one is demanding an enhanced constant cross-section 
\cite{Sood04} in QMD model. On the other hand,
it is also observed that the reduced isospin dependent cross-section is valid for the soft as well as
for soft momentum dependent equation of state \cite{Zhang99,Gautam10}
 within IQMD or IBUU model. The systematic concept of enhanced 
and reduced isospin dependent cross-sections in the presence of hard equation of state is missing throughout the 
literature. Moreover, the effect of isospin dependent cross-sections are studied on the limited systems 
experimentally as well as theoretically \cite{Pak97,Li96,Leiwen98,Gautam10,Hartnack98}. We plan to 
study the effect of enhanced (30 \% of $\sigma_{NN}$) as well as reduced (30 \% of $\sigma_{NN}$) 
isospin dependent cross-sections in the presence of hard equation of state on the systems for which 
the experimental finding energy is predicted in the literature and then will compare the results
with experimental findings. For this study, we will employ isospin dependent Quantum Molecular Dynamics(IQMD) model
which is discussed in sec. II. The results are discussed in sec. III, followed by conclusion in sec. IV.  
\section{ISOSPIN-Dependent QUANTUM MOLECULAR DYNAMICS (IQMD) MODEL}
The isospin-dependent quantum molecular dynamics (IQMD) \cite{Hartnack98} model treats different charge states of
nucleons, deltas and pions
explicitly \cite{Hartnack98}, as inherited from the VUU model \cite{Krus85}. 
The IQMD model has been used successfully
for the analysis of large number of observables from low to relativistic energies \cite{Gautam10}. 
The isospin degree of
freedom enters into the calculations via symmetry potential, cross-sections and
Coulomb interactions \cite{Krus85}.
The details about the elastic and inelastic cross-sections
for proton-proton and neutron-neutron collisions can be found in Ref. \cite{Hartnack98}. These cross-sections follow the 
data published by particle data group (PDG) for proton-neutron and proton-proton scattering \cite{Amsler08}. 
In this model, baryons are represented by Gaussian-shaped density distributions
\begin{equation}
f_i(\vec{r},\vec{p},t) = \frac{1}{\pi^2\hbar^2}\cdot e^{-(\vec{r}-\vec{r_i}(t))^{2}\frac{1}{2L}}\cdot e^{-(\vec{p}-\vec{p_i}(t))^{2}\frac{2L}{\hbar^2}}.
\end{equation}
Where L is the Gaussian Width.
As mentioned in Ref. \cite{Hartnack98}, in IQMD the value of Gaussian width L depends on the size of the system. 
This system size dependence of L in IQMD has been introduced in order to obtain the maximum stability of the
nucleonic density profile. Therefore, in the present study, by checking the stability, 
we have  taken the value from 0.5L to L. Its earlier version QMD has
been very successful in explaining the multifragmentation \cite{Gossiaux97},
temprature and density \cite{Khoa92}, flow \cite{Kumar98},
multifragments \cite{Singh00} and particle production \cite{Huang93}. \\ 
  Nucleons are initialized in a sphere with radius $R= 1.12 A^{1/3}$ fm, 
in accordance with the liquid drop model. 
Each nucleon occupies a volume of $h^3$, so that phase space is uniformly filled. The initial momenta 
are randomly chosen between 0 and Fermi momentum($\vec P_{F}$). The nucleons of target and projectile
interact via two and three-body Skyrme forces, Yukawa potential, Coloumb interactions. 
In addition to the use of explicit charge states of all baryons and mesons a symmetry potential between 
protons and neutrons corresponding to the Bethe- Weizsacker mass formula has been included.\\
The hadrons propagate using Hamilton equations of motion:
\begin{equation}
\frac{d\vec{r_i}}{dt}~=~\frac{d\it{\langle~H~\rangle}}{d{p_i}}~~;~~\frac{d\vec{p_i}}{dt}~=~-\frac{d\it{\langle~H~\rangle}}
{d{r_i}},
\end{equation}
with
\begin{eqnarray}
\langle~H~\rangle&=&\langle~T~\rangle+\langle~V~\rangle\nonumber\\
&=&\sum_{i}\frac{p_i^2}{2m_i}+
\sum_i \sum_{j > i}\int f_{i}(\vec{r},\vec{p},t)V^{\it ij}({\vec{r}^\prime,\vec{r}})\nonumber\\
& &\times f_j(\vec{r}^\prime,\vec{p}^\prime,t)d\vec{r}d\vec{r}^\prime d\vec{p}d\vec{p}^\prime .
\end{eqnarray}
 The baryon-baryon potential $V^{ij}$, in the above relation, reads as:
\begin{eqnarray}
V^{ij}(\vec{r}^\prime -\vec{r})&=&V^{ij}_{Skyrme}+V^{ij}_{Yukawa}+V^{ij}_{Coul}+V^{ij}_{sym}\nonumber\\
&=&\left(t_{1}\delta(\vec{r}^\prime -\vec{r})+t_{2}\delta(\vec{r}^\prime -\vec{r})\rho^{\gamma-1}
\left(\frac{\vec{r}^\prime +\vec{r}}{2}\right)\right)\nonumber\\
& & +~t_{3}\frac{exp(|\vec{r}^\prime-\vec{r}|/\mu)}{(|\vec{r}^\prime-\vec{r}|/\mu)}~+~\frac{Z_{i}Z_{j}e^{2}}{|\vec{r}^\prime -\vec{r}|}\nonumber\\
& &+t_{6}\frac{1}{\varrho_0}T_3^{i}T_3^{j}\delta(\vec{r_i}^\prime -\vec{r_j}).
\label{s1}
\end{eqnarray}
Here $Z_i$ and $Z_j$ denote the charges of $i^{th}$ and $j^{th}$ baryon, and $T_3^i$, $T_3^j$ are their respective $T_3$
components (i.e. 1/2 for protons and -1/2 for neutrons). Meson potential consists of Coulomb interactions only.
The parameters $\mu$ and $t_1,.....,t_6$ are adjusted to the real part of the nucleonic optical potential. For the density
dependence of nucleon optical potential, standard Skyrme-type parametrization is employed.
The choice of equation of state (or compressibility) is still controversial one. Many studies
advocate softer matter, whereas, much more believe the matter to be harder in
nature \cite{Krus85,Mage00}. 
For the present analysis, a hard (H) equation of state,
has been employed along with standard energy dependent cross-sections. Note that the relativistic effects are 
neglisible at these enegies \cite{Lehmann93}. \\
\section{Results and Discussion}
We study the directed flow using a stiff equation of state along with enhanced and reduced isospin dependent cross-sections
($\sigma$= 0.7 to 1.3 $\sigma_{NN}$), by simulating various reactions. The time 
evolution of reaction is follow upto 200 fm/c. This is the time at which transverse in-plane flow saturates
for lighter as well as for heavier systems. For this study, the reactions of $^{12}C_{6}~+~^{12}C_{6}$
($\hat{b}=0.4$, L=0.5L) where L=8.66 $fm^2$, $^{20}Ne_{10}~+~^{27}Al_{11}$ ($\hat{b}=0.4$, L=0.5L),
$^{40}Ar_{18}~+~^{45}Sc_{21}$ ($\hat{b}=0.4$, L=0.5L), $^{40}Ar_{18}~+~^{51}V_{23}$ ($\hat{b}=0.3$, L=0.5L),
$^{86}Kr_{36}~+~^{93}Nb_{41}$ ($\hat{b}=0.4$, L=0.6L), $^{64}Zn_{30}~+~^{58}Ni_{28}$ ($\hat{b}=2fm$, L=0.6L),
$^{93}Nb_{41}~+~^{93}Nb_{41}$ ($\hat{b}=0.3$, L=0.7L), $^{129}Xe_{54}~+~^{118}Sn_{50}$ ($\hat{b}=0-3fm$, L=0.7L),
$^{139}La_{57}~+~^{139}La_{57}$($\hat{b}=0.3$, L=0.8L), $^{197}Au_{79}~+~^{197}Au_{79}$($\hat{b}=2.5fm$, L=L)
are simulated. The choice of impact parameter is guided by the experimentally extracted information 
\cite{Westfall93,Pak96,Pak97,Sullivain90,Cussol02}. The above reactions were simulated between 45 and 
200 MeV/nucleon using the hard equation of state along with different isospin dependent cross-sections. 
We have attempted to fit the reduced isospin dependent cross-sections in the presence of stiff equation of state with 
experimental findings, as is performed in the literature with soft equation of state with and without momentum 
dependent interactions.\\
There are two methods in the literature used to find the balance energy \cite{Sood04}. In the first case, 
the balance energy is extracted from the $\langle P_x/A \rangle$ Plots, where $\langle P_x/A \rangle$ is 
to plotted as a function of rapidity distribution $Y_{c.m.}$/$Y_{beam}$, which is given as.
\begin{equation}
   Y(i)=\frac{1}{2}~ln\frac{E(i)+P_z(i)}{E(i)-P_z(i)}
\end{equation}
where E(i) and $P_z(i)$ are respectively, the total energy and longitudinal momentum of $i^{th}$ particle.
Naturally, the energy at which this flow passes through zero is called balance energy. The second method 
is to study the incident energy dependence of the directed transverse in-plane flow 
$\langle P_{x}^{dir} \rangle$, which is defined as \cite{Sood04}
\begin{equation}
  \langle P_x^{dir}\rangle=\frac{1}{A}\sum_{i}^A sign\{Y(i)\}P_{x}(i)
\end{equation}
where Y(i) is the rapidity distribution as discussed above and $P_{x}(i)$ is the transverse momentum of the $i^{th}$
particle in x-direction. This $\langle P_x^{dir}\rangle$ is defined over entire rapidity region and therefore expected to
present an easier way of measuring the in-plane flow rather than complicated $\langle P_x/A \rangle$ plots.
In the present study, we have tried to study the effect of isospin dependent cross-sections on the flow or 
alternatively on the balance energy by using both of the parameter and then the detailed study is 
extended with later one.\\
\begin{figure}
\includegraphics[width=0.5\textwidth]{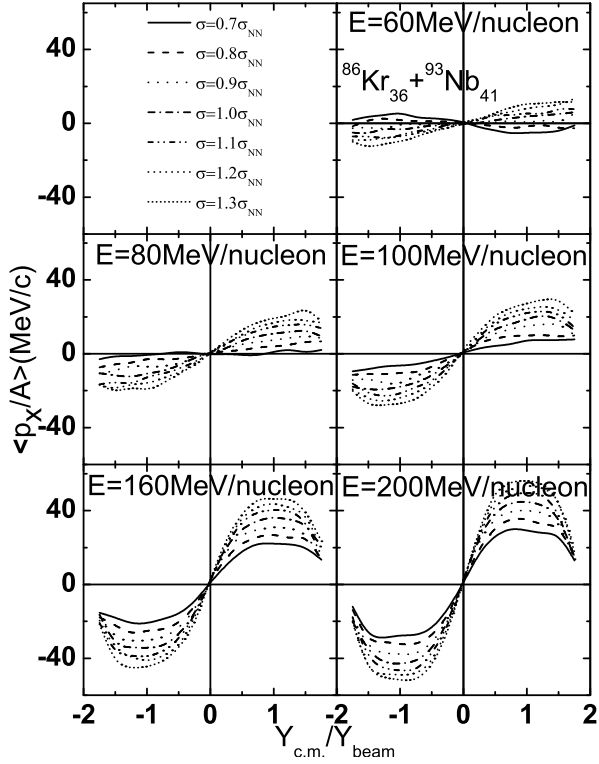}
\caption{\label{fig:1}The averaged $\langle P_x/A \rangle$ as function of the rapidity distribution. Here we 
display the result for Kr+Nb system at different incident energies and different isospin-dependent 
cross-sections.}
\end{figure}
                                                                                                                             
\begin{figure}
\includegraphics[width=0.5\textwidth]{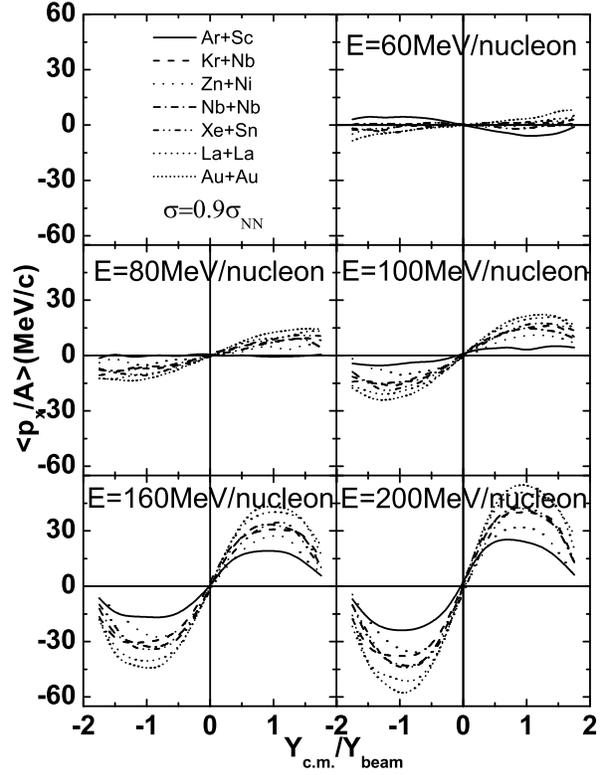}
\caption{\label{fig:2} As in \ref {fig:1}, but for different system and at particular cross-section 
$\sigma=0.9\sigma_{NN}$. Diffrent panels are at diffrent incident energies.} 
\end{figure}

\begin{figure}
\includegraphics[width=0.5\textwidth]{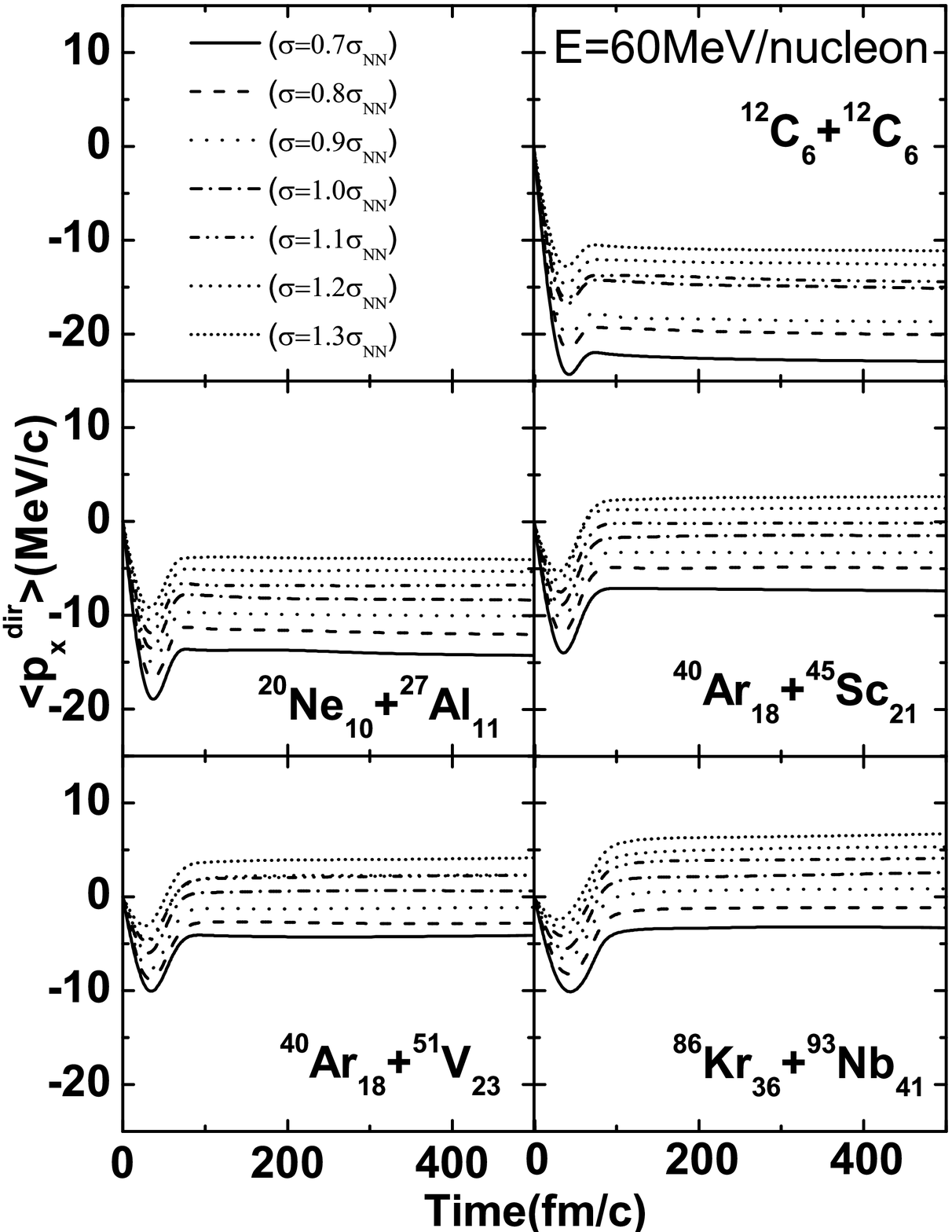}
\caption{\label{fig:3} Time evolution of $\langle P_x^{dir}\rangle$ for different systems at E=60 MeV/nucleon.
The different lines in figure representing the directed fow at different cross-sections.}
\end{figure}
                                                                                                                             
\begin{figure}
\includegraphics[width=0.5\textwidth]{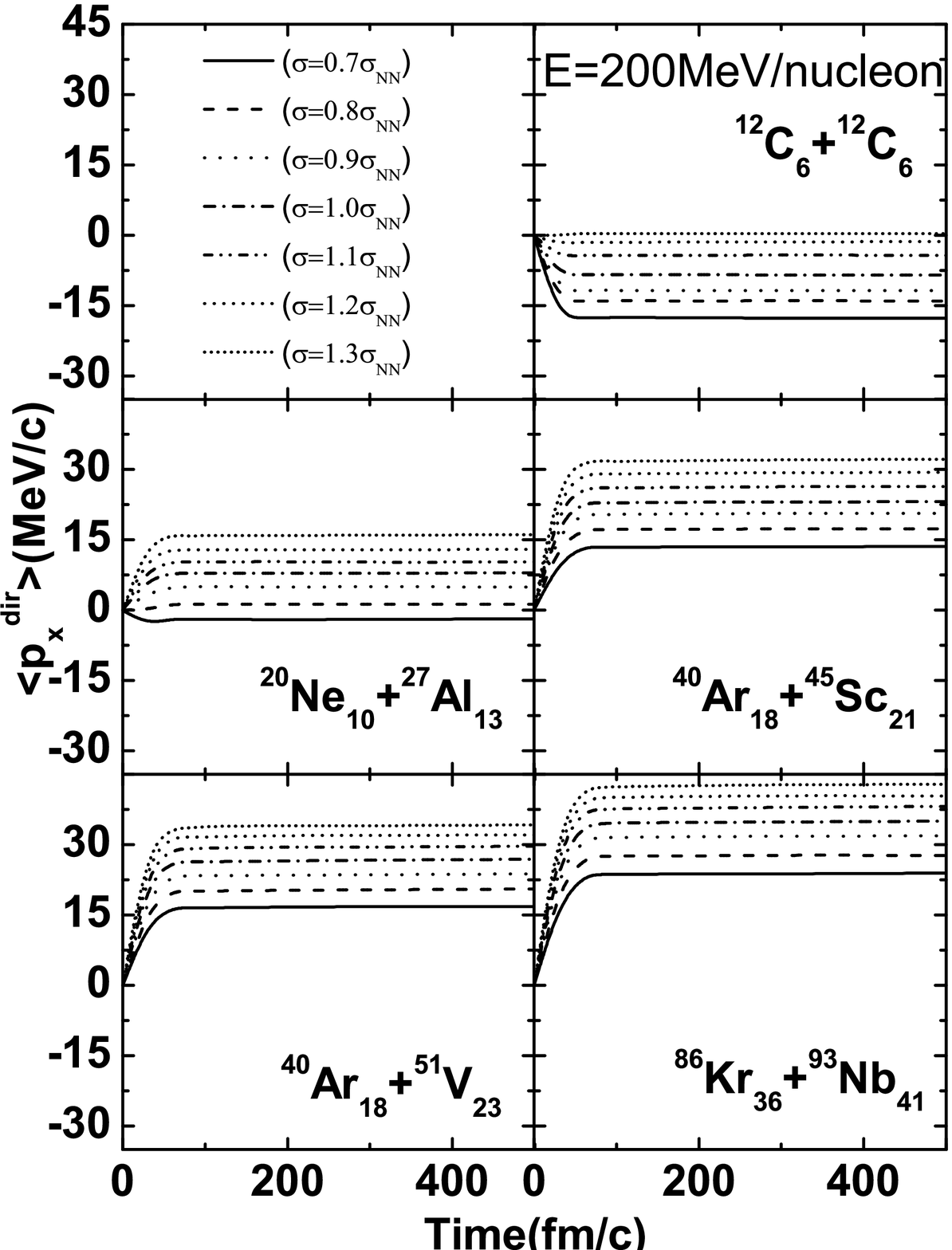}
\caption{\label{fig:4} Time evolution of $\langle P_x^{dir}\rangle$ for different systems at E=200MeV/nucleon.
The different lines in figure representing the directed fow at different cross-sections.}
\end{figure}

\begin{figure}
\includegraphics[width=0.5\textwidth]{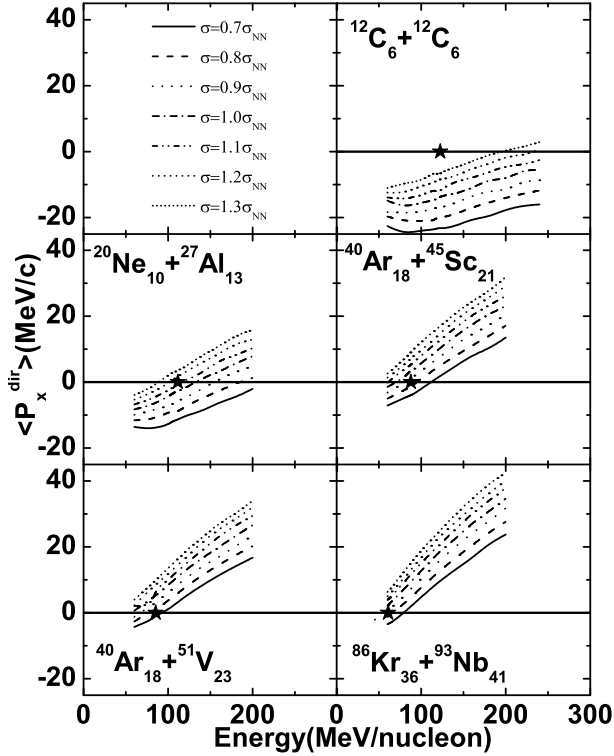}
\caption{\label{fig:5} Energy dependence of the directed nuclear flow $\langle P_x^{dir}\rangle$
for different systems. The lines have same meaning as that in fig 3 and fig 4.}
\end{figure}
                                                                                                                             
\begin{figure}
\includegraphics[width=0.5\textwidth]{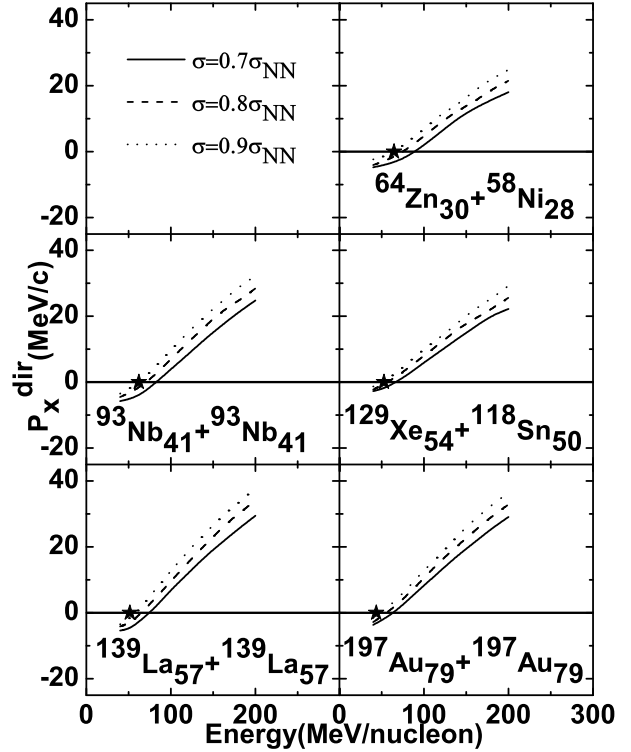}
\caption{\label{fig:6} same as in fig 5, but for different systems.}
\end{figure}
                                                                                                                             
\begin{figure}
\includegraphics[width=0.5\textwidth]{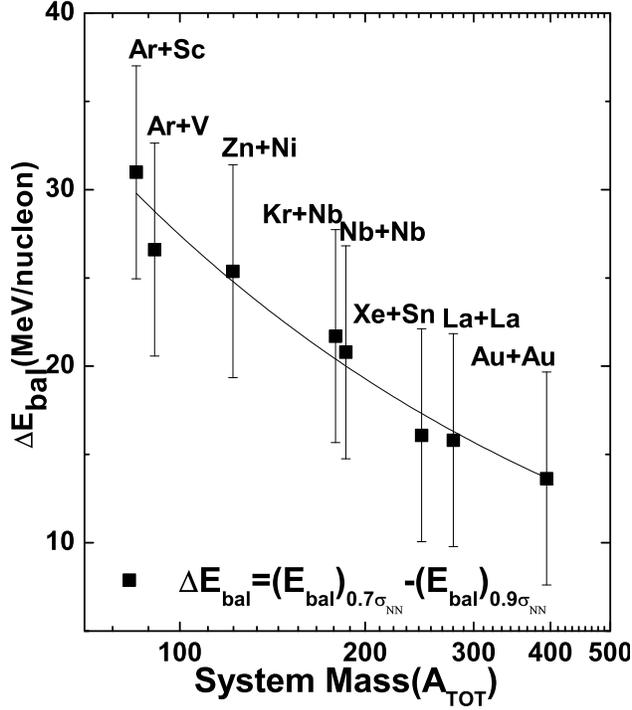}
\caption{\label{fig:7} The shift in  balance energy due to cross section as a function of combined mass of the
system.}
\end{figure}

\begin{figure}
\includegraphics[width=0.5\textwidth]{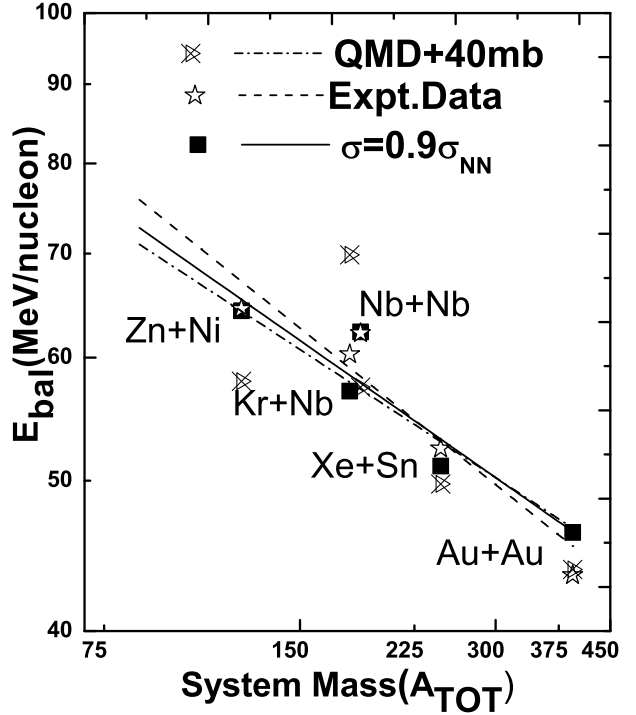}
\caption{\label{fig:8} Balance energy as a function of combined mass of the system. The experimental
points are represented with stars, QMD+40mb with crossed triangle and present with solid square. }
\end{figure}

\begin{figure}
\includegraphics[width=0.5\textwidth]{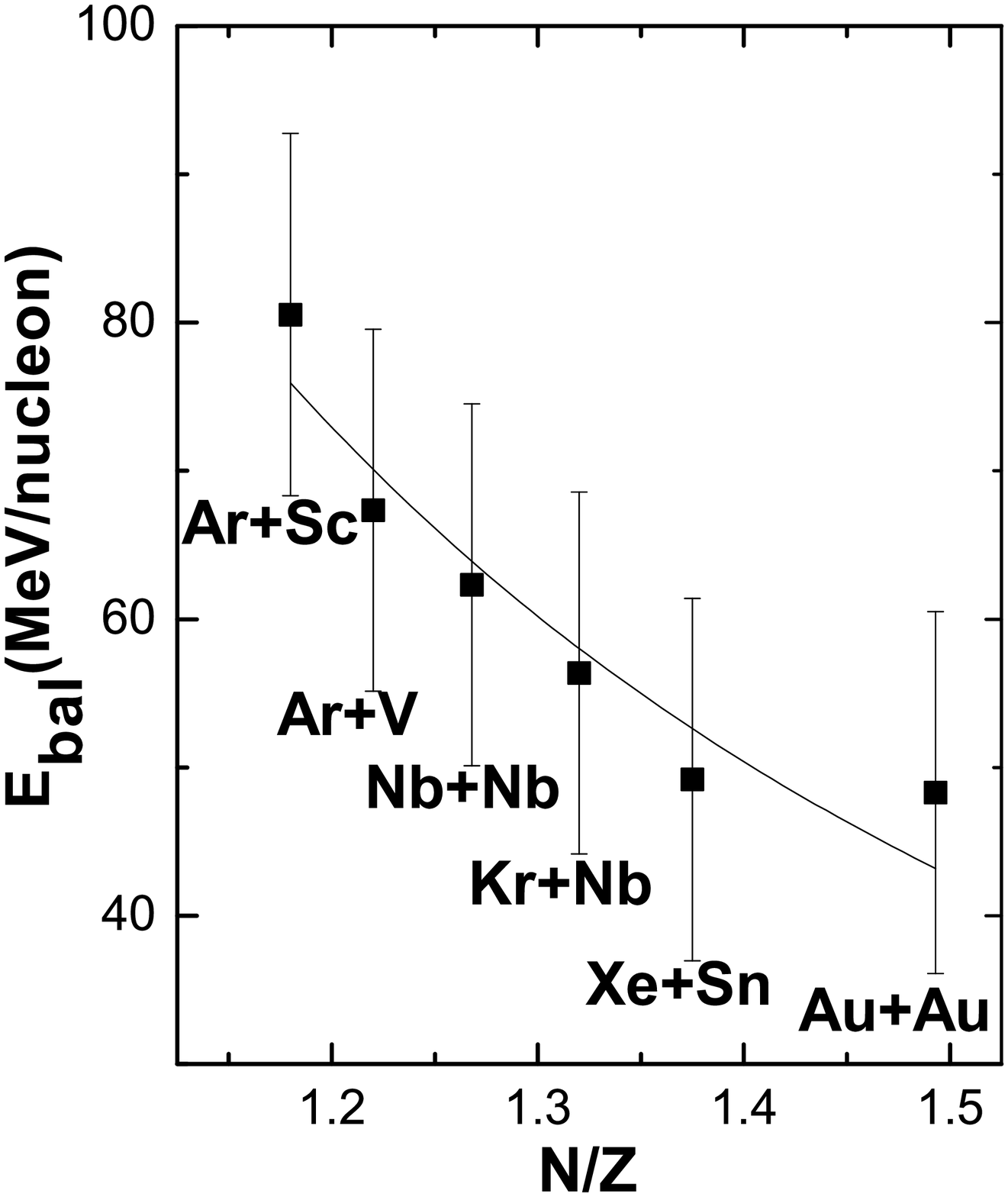}
\caption{\label{fig:9} N/Z dependence of balance energy at $\sigma=0.9\sigma_{NN}$ The curve is parametrized
with power law. }
\end{figure}

In Fig. 1, we display the change in the transverse momentum $\langle P_x/A \rangle$ as a function of the rapidity 
distribution at different incident energies from 60 to 200 MeV/nucleon for $^{86}Kr_{36}~+~^{93}Nb_{41}$ systems. 
The different lines in the figure are showing the variation with different cross-section values. From the figure, we see 
that slope becomes less negative or more positive with increase in the incident energy. On the other hand,
with reduction in the cross-section ($\sigma_{NN}$), slope is getting more negative or less positive, while,
becoming more positive with enhanced cross-section. This indicates that we see a change in the slope with incident 
energy and reduced isospin dependent cross-section. 
  The figure is indicating two values of balance energy i.e. E=80MeV/nucleon (at $\sigma=0.7\sigma_{NN}$) 
and around E=60MeV/nucleon (at $\sigma=0.9\sigma_{NN}$) for $^{86}Kr_{36}~+~^{93}Nb_{41}$ system. As the 
experimental balance energy for $^{86}Kr_{36}~+~^{93}Nb_{41}$ is in the range of 55-60 MeV/nucleon, So one is
 expecting to follow the whole dynamics at $\sigma=0.9\sigma_{NN}$.\\
Further, the detailed analysis of rapidity distribution of transverse momentum $\langle P_x/A \rangle$ 
for different systems at different energies with $\sigma=0.9\sigma_{NN}$ is displayed in Fig. 2.
The slope is becoming more positive or less negative with increase in the composite mass of system, indicating
that lighter systems remain in the environment of mean field compared to NN collisions at any 
given incident energy. The contribution of the mean field the and collisions is discussed in detail in Ref. \cite{Sood04}.
One also notice that a higher incident energy is needed in lighter cases to balance the attractive and repulsive forces.
This energy is supposed to decrease with increase in the system mass. Similar findings are also published by Puri and
co-worker\cite{Sood04}. Note that their study did not take isospin effects into consideration.\\
                                                                                                                                                Figures 3 and 4 are displaying the time evolution of the second parameter $\langle P_x^{dir}\rangle$
below (60MeV/nucleon) and above (200MeV/nucleon) the experimental balance energy, respectively. The results 
in the figure are displayed for five different systems and at reduced as well as enhanced isospin 
dependent cross-sections. The figures are indicating the similar scenerio with $\langle P_x^{dir}\rangle$ 
as is depicted with $\langle P_x/A \rangle$. Below the balance energy (in Fig. 3), the directed in-plane flow is negative
during the initial phase of reaction for all the systems under consideration. This becomes positive at sufficient 
high incident energy say E=200 MeV/nucleon (in Fig. 4). These results shows that interaction among nucleons 
are attractive during the initial phase of the reaction, which turns out to be repulsive with increase in the
incident energy. These interactions remain either attractive or repulsive throughout the time evolution depends on
the incident energy, isospin dependent cross-sections as well as composite mass of the system. It is clear from the figure 
that directed flow is becoming more positive or less negative with incident energy, isospin dependent cross-sections
as well as with size of system. There is a sharp transition for each system from negative to positive directed flow 
at a particular cross-section. This particular transition is not possible for $^{12}C_{6}~+~^{12}C_{6}$ system indicates the 
requirement of other variable like momentum dependent interaction as well as enhancement in cross-section
more then 30 \%. If one compares the figs. 1-4, the same physics of balance energy elaborates with the
$\langle P_x/A \rangle$ as well as with $\langle P_x^{dir}\rangle$. Out of these, as discussed earlier,
 $\langle P_x^{dir}\rangle$ is the simple and more useful quantity, because it is summed over entire rapidity distribution,
that is why, $\langle P_x^{dir}\rangle$ is elaborated in detail for the further study.\\
To study the influence of reduced as well as enhanced cross-sections on directed flow 
$\langle P_x^{dir}\rangle$ or alternatively on
the balance energy, in Figs 5 and 6, incident energy dependence of directed flow is displayed for different 
systems. The different lines in figure represent the variation with different cross-sections.                                  The studies with enhanced and reduced cross-sections are also available in the 
literature \cite{Sood04,Zhang99,Gautam10}. The experimental data are represented by stars. The directed flow 
goes from negative to positive value with increase in the incident energy. This is the general trend and 
is explained many times in the litrature by taking the concept of mean field and NN cross-sections. On the other hand,
the role of different cross-sections is consistent through the present mass range. By finding
the evidence of reduced cross-sections from Fig. 5, the results are displayed between (0.7-0.9 $\sigma_{NN}$)values
in Fig. 6. The enhanced cross-section (1.3 $\sigma_{NN}$) gives more positive value 
followed by the cugnon cross-section ($\sigma_{NN}$) towards the reduced 
cross-section (0.7 $\sigma_{NN}$). In other words, with increase in the cross-section value from 
(0.7$\sigma_{NN}$-1.3$\sigma_{NN}$), the directed flow is becoming more positive or less negative. 
This is due to the reason that with increase in the cross-section value, probability of reaction to take place increases
that further results increase in the NN collisions and hence more positive value of the directed flow. This is 
resulting decrease in the balance energy. The balance energy is also found to decrease with 
increase in the composite mass of the system. This is due to dominance of Coulomb repulsion with an increase in the 
composite mass of system. Except for some lighter systems, the cross-section $\sigma=0.9\sigma_{NN}$ is found to  
explain the experimental balance energy nicely. Similar parametrization was also performed by Sood {et al.} \cite{Sood04}
within QMD model. They also found that enhanced cross-section ($\sigma=40mb$) can best explain the data. In contrary,
calculation in IQMD model demand reduced value of cross-sections.
The difference is due to the additional effect of isospin dependent cross-sections in IQMD model \cite{Hartnack98}, 
which were absent in QMD model. As in QMD model, the strength of nn, pp, pp cross-section is taken equal, while in IQMD,
$\sigma_{np}~=~3\sigma_{pp} \approx 3\sigma_{nn}$ \cite{Hartnack98}. Due to the different strength 
of np, pp, nn cross-section in IQMD, additional repulsion is produced compared to QMD model. This addition 
repulsion will force the directed flow to take earlier transition from negative to positive value and hence 
will lower the balance energy in IQMD model as compared to QMD for same cross-section value. 
That is why, the balance energy that was obtained with QMD at $\sigma=40mb$
are at $\sigma=0.9\sigma_{NN}$ in IQMD model. This is first ever 
parametrization of balance energy with hard equation of state
in the presence of reduced cross-sections with experimental available balance energy.\\
 By taking the Ref. \cite{Sood04}
into account, which depicts, that for heavier colliding nuclei $E_{bal}$ is independent of the cross-section
one is choosing, we have plotted in Fig. 7 the 
$\Delta E_{bal}$~=~$(E_{bal})_{0.7\sigma_{NN}}-(E_{bal})_{0.9\sigma_{NN}}$
with composite mass of system. Our findings are also supporting the findings of Ref. \cite{Westfall93,Li99}. $\Delta E_{bal}$
is maximum for lighter systems and it goes on decreasing with system mass. It is also indicating 
theoretically that balance energy is almost independent of the nucleon-nucleon cross-section 
for the heavier system such as Au+Au, U+U etc.\\
 In Fig. 8, we display the energy of vanishing flow or balance energy($E_{bal}$) as a function of 
composite mass of system that ranges from $^{40}Ar_{18}+^{45}Sc_{21}$ to $^{197}Au_{79}+^{197}Au_{79}$. 
In this figure, $E_{bal}$ is showed for the 
experimental data (open stars), QMD+40mb (crossed triangle) and IQMD+0.9 $\sigma_{NN}$ (solid squares). All the
curves are fitted with power law of the form $C(A_{TOT})^\tau$. The experimental data are fitted by 
$\tau=-0.33 \pm 0.06 $ The balance energy is found to decrease with the composite 
mass of the system, which is a well known 
trend discussed many times in literature \cite{Sood04}. The difference is in the $\tau$ values 
obtained by different theoretical model. The BUU model report $\tau$ between 
$-0.28\leq \tau^{th} \leq -0.32$. In another study \cite{Westfall93}
again with BUU model $\tau^{th}=-0.41 \pm 0.03$. The present calculation depicts the 
$\tau$ value $(-0.29 \pm 0.06)$, which is close to the experimental $\tau$ value $(-0.33 \pm 0.06)$ as 
compared to QMD+40mb calculation having $\tau$ value $(-0.27 \pm 0.17)$. In other words, the present IQMD model with a 
stiff equation of state along with $\sigma=0.9\sigma_{NN}$ can explain the data much better than any other 
theoretical calculations. The $\sigma=0.9\sigma_{NN}$ explains the data for all nuclei, except for some 
lighter nuclei. The lighter nuclei, when checked out, demand for an enhanced cross-sections \cite{Sood04,Kumar10}
along with momentum dependent interactions \cite{Gossiaux97}. Our calculations about the strength of reduced NN 
cross-section is in agreement with earlier calculation, where disappearance of transverse in-plane flow \cite{Gautam10}
as well as elliptical flow is parametrized with experimental data \cite{Zhang99}.\\
We have also tried to fit the balance energy in terms of other parameter such as the neutron to proton ratio of
colliding nuclei. This attempt is shown in Fig. 9, where balance energy is plotted as a function of N/Z. The $E_{bal}$
is parametrized with power law of the form $(N/Z)^\tau$. The $\tau$ value in N/Z dependence is $-2.39 \pm 0.40$,
 while in $A_{TOT}$ dependence in Fig. 8 is $-0.29 \pm 0.06$. The $\tau$ value in this case is larger  
compared to the mass dependence. The difference in the slopes may be due to different charge to mass ratio in 
heavier colliding nuclei.\\
\section{Conclusion} 
By using the IQMD model, we have studied the effect of reduced as well as enhanced isospin dependent
cross-sections on the directed flow and balance energy. A large number of reactions were studied having mass
range from 24 to 394, where experimental balance energy is available. Our calculation with stiff equation of 
state and reduced cross-section ($\sigma=0.9\sigma_{NN}$) are in good agreement with the experimental findings,
except for $^{12}C_{6}+^{12}C_{6}$. The dependence of isospin dependent cross-sections get weakens with 
increase in the size of system. The balance energy is parametrized with N/Z ratio in terms of 
power law, which is to be quite similar with the parametrization of composite mass of system, but the 
$\tau$ values are different in both of the cases. One could try the balance energy prediction with 
enhanced isospin dependent cross-section in the presence of momentum dependent interaction for
$^{12}C_{6}+^{12}C_{6}$, which is earlier studied by Sood {et al.} \cite{Sood04} by using QMD in the presence of 
momentum dependent interactions.\\
\begin{acknowledgments}
This work has been supported by the Grant no. 03(1062)06/ EMR-II, from the Council of Scientific and
Industrial Research (CSIR) New Delhi, Govt. of India.\\
\end{acknowledgments}

\end{document}